\documentclass[twoside]{LCWS11}
\usepackage[latin1]{inputenc}
\usepackage[dvips]{graphicx,epsfig,color}
\usepackage{wrapfig,rotating}
\usepackage{amssymb,amsmath,array}
\usepackage{cite}
\newcommand{\GeV}{\ensuremath{\enspace\mathrm{GeV}}}
\newcommand{\Pe}{P_{e^{-}}}
\newcommand{\Pp}{P_{e^{+}}}
\newcommand{\ep}{e^{+}e^{-}}
\newcommand{\nng}{\nu\nu\gamma}
\newcommand{\nngN}{\nu\nu\gamma (N)\gamma}
\newcommand{\lum}{\mathcal{L}}
\newcommand{\fb}{\ensuremath{\enspace\mathrm{fb}}}
\newcommand{\nt}{\tilde{\chi}^0}
\begin{document}
\title{
Measurement of radiative neutralino production} 
\author{Christoph~Bartels$^1$, Olaf~Kittel$^2$\thanks{Speaker},
Ulrich~Langenfeld$^3$\thanks{U. L. has been supported by funding from the research 
training group GRK 1147 of the Deutsche Forschungsgemeinschaft and 
partially by the Helmholtz alliance 'Physics at the Terascale. },
Jenny~List$^1$
\vspace{.3cm}\\
1- DESY, Notkestrasse 85, D-22607 Hamburg, Germany
\vspace{.1cm}\\
2- Departamento de F\'isica Te\'orica y del Cosmos and CAFPE, \\
Universidad de Granada, E-18071 Granada, Spain
\vspace{.1cm}\\
3- Institut f\"ur Theoretische Physik und Astronomie, \\
Universit\"at W\"urzburg, Am Hubland, D-97074 W\"urzburg, Germany
}
\maketitle
\begin{abstract}
We perform the first experimental study with full detector simulation
for the radiative production of neutralinos at the linear collider,
at $\sqrt{s} =500\GeV$ and realistic beam polarizations.
We consider all relevant backgrounds, like the Standard Model background 
from radiative neutrino production.
The longitudinal polarized beams enhance the signal and simultaneously
reduce the background, such that statistical errors are significantly reduced.
We find that the photon spectrum from the signal process can be well isolated.
The neutralino mass and the cross section can be 
measured at a few per-cent level, with the largest systematic
uncertainties from the measurement of the beam polarization
and the beam energy spectrum.
\end{abstract}

\section{Introduction}

\begin{figure}[t!]
\setlength{\unitlength}{1cm}
 \begin{picture}(20,6)(-2,0)
 \scalebox{0.35}{\includegraphics{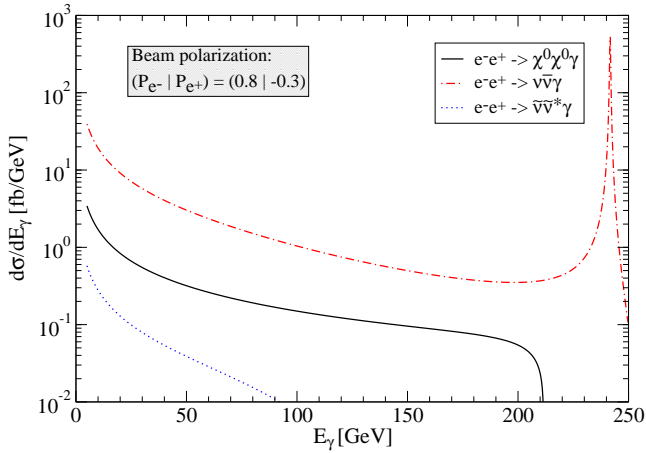}}
\vspace*{3.0cm}
\end{picture}
\caption{
     Photon energy distributions for 
      $e^+e^- \to \tilde\chi^0_1 \tilde\chi^0_1\gamma$ (black, solid),
       $e^+e^- \to \nu\bar\nu\gamma$ (red dot-dashed), and
      $e^+ e^- \to \tilde\nu\tilde\nu^\ast\gamma$ (blue, dotted), 
      at the LC for $\sqrt{s}=500$~GeV, ${\mathcal{L}}=500~{\rm fb}^{-1}$,
      and $(P_{e^-},P_{e^+})=(0.8,-0.3)$,
      for scenario Eq.~(\ref{Eq:scenario}), 
      with regularization cut $|\cos\theta_\gamma| \le 0.99$~\cite{Dreiner:2006sb}.
}
\label{plotEdist}
\end{figure}

The Minimal Supersymmetric Standard Model (MSSM) is a promising
extension of the Standard Model of particle physics (SM)~\cite{Haber:1984rc}.
At a future Linear Collider (LC),
the masses, decay widths, couplings,
and spins of the new SUSY particles can be measured with high
precision~\cite{AguilarSaavedra:2005pw}. 
In particular, the lightest electroweak states like pairs of neutralinos, 
charginos, and sleptons, can be studied in the initial stage of the LC, with a 
center-of-mass energy $\sqrt s = 500$~GeV, and a luminosity of order 
${\mathcal L}=500$~fb$^{-1}$. The lightest visible SUSY state is a pair of 
radiatively produced 
neutralinos~\cite{Choi:1999bs,Baer:2001ia,Dreiner:2006sb,Dreiner:2007vm}
\begin{equation}
e^+ + e^-\to\tilde\chi_1^0 + \tilde\chi_1^0 + \gamma. 
\label{neut-pairs}
\end{equation}
The signal is a single high energetic photon, radiated off the incoming 
beams or off the exchanged selectrons, and missing energy from the neutralinos.
The main irreducible Standard Model background is photons from radiatively  
produced neutrinos $e^+e^- \to \nu \bar\nu \gamma$, see  Fig.~\ref{plotEdist}.
Indeed, due to this large background, radiative neutralino production cannot 
be observed at LEP~\cite{Dreiner:2006sb}, not even for light or even  
massless neutralinos~\cite{Dreiner:2009ic}.

\medskip

\begin{figure}
\setlength{\unitlength}{1cm}
 \begin{picture}(20,10)(0.7,0)
 \put(2.5,9.5){\fbox{$\sigma(e^+e^- \to \tilde\chi^0_1\tilde\chi^ 0_1\gamma)$ in fb}}
 \put(0.3,2.5){\includegraphics{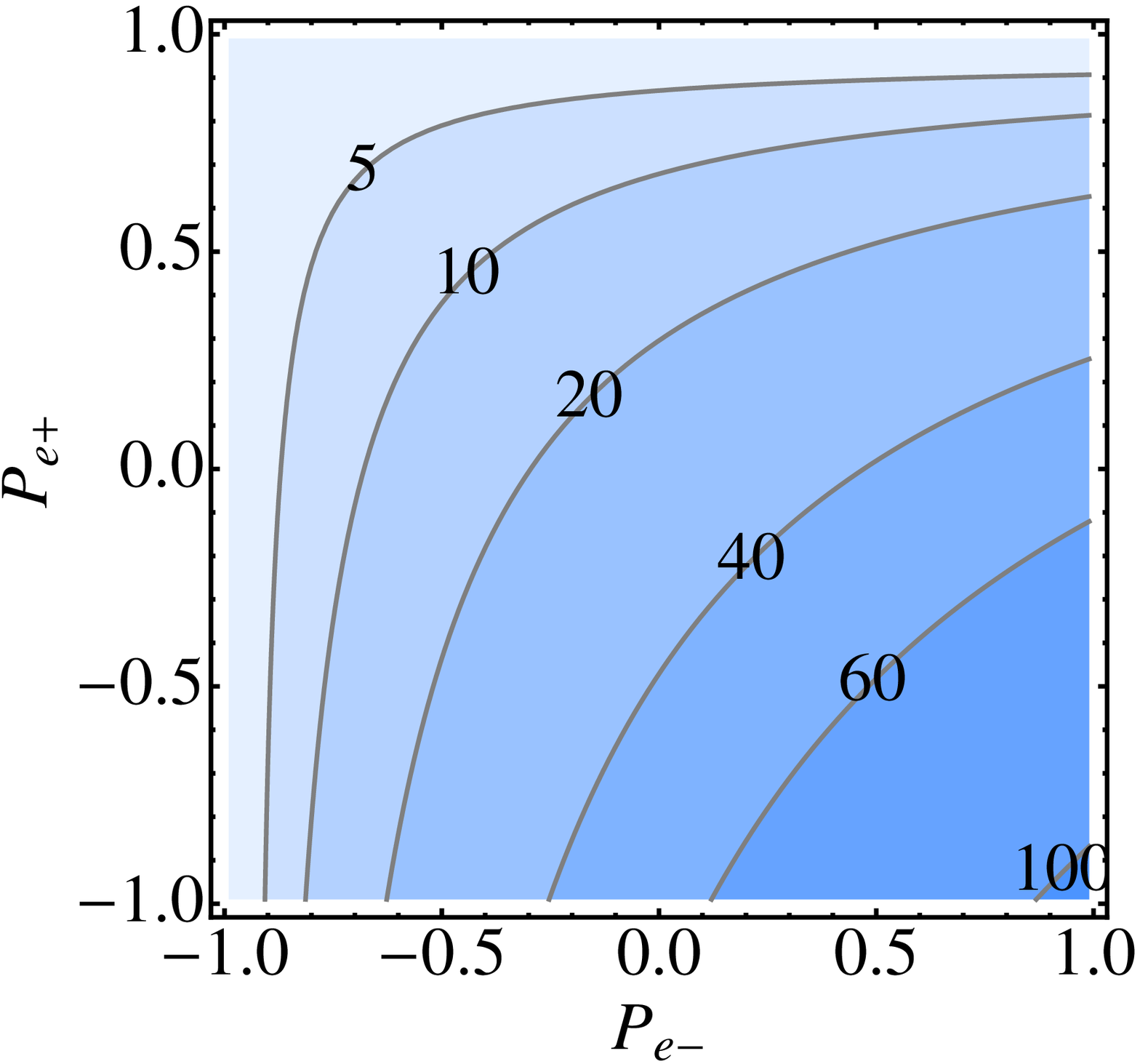}}
  \put(7.7,2.5){\includegraphics{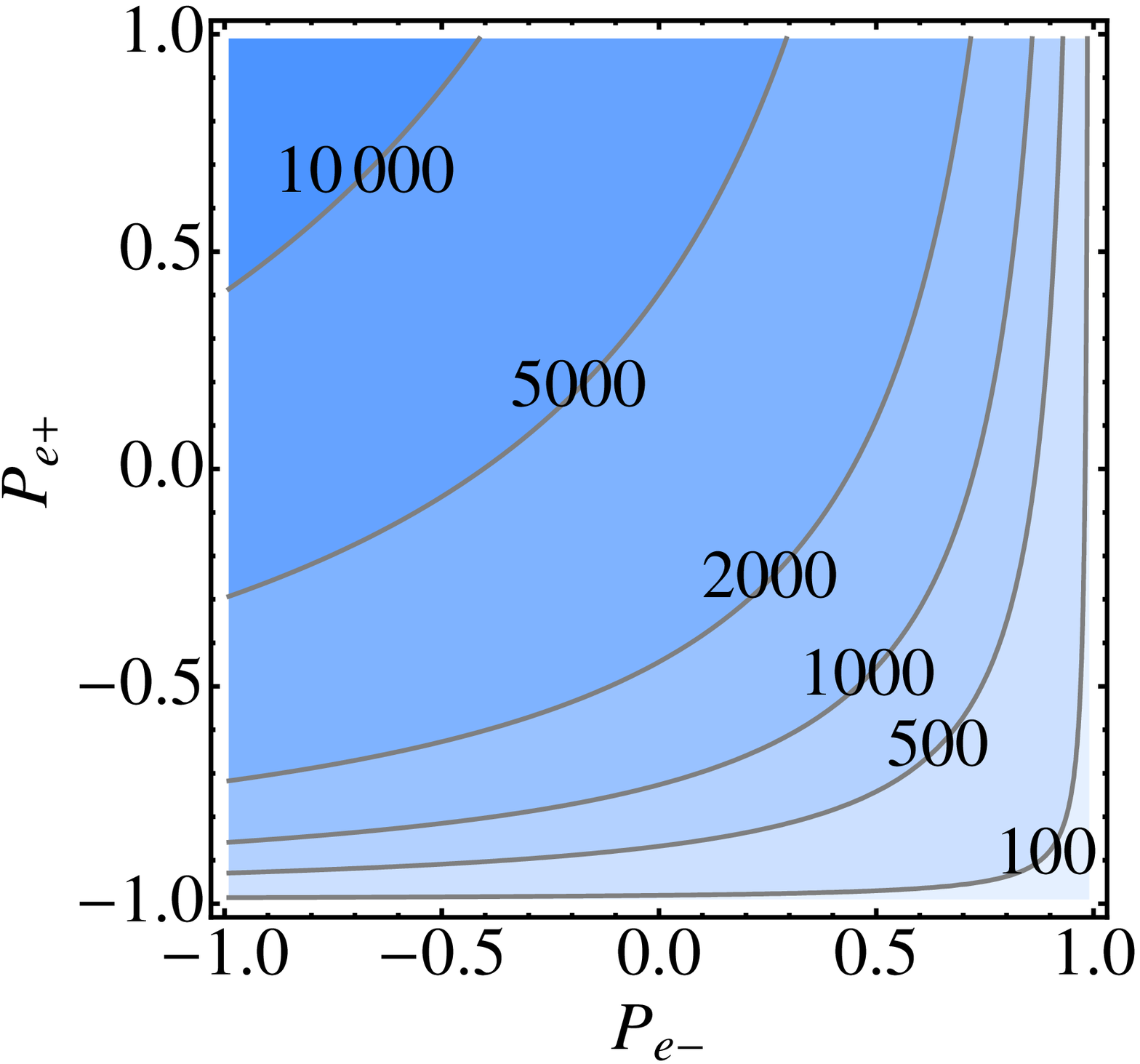}}
  \put(10.,9.5){\fbox{$\sigma_{\rm B}(e^+e^-\to\nu\bar\nu \gamma)$ in fb }}
\end{picture}
\vspace*{-3.0cm}
\caption{%
        Contour lines of the tree-level cross section~\cite{Dreiner:2006sb,Dreiner:2007vm}
 for radiative neutralino production,
         $e^+e^- \to \tilde\chi^0_1\tilde\chi^0_1\gamma$ (left) and
        radiative neutrino production  $e^+e^- \to \nu\bar\nu\gamma$ (right),
         at the LC for $\sqrt{s}=500$~GeV and ${\mathcal{L}}=500~{\rm fb}^{-1}$, 
        for our benchmark scenario, Eq.~(\ref{Eq:scenario}). We applied  
        regularization cuts on the photon angle
         $|\cos\theta_\gamma| \le 0.99$, and energy $5 \le E_\gamma[{\rm GeV}]\le 212$.
        \label{varBeamPol}}
\end{figure}

The discovery potential for the radiative production of light neutralinos
is much better for the LC, since it provides  high luminosity and
the option of polarized beams~\cite{MoortgatPick:2005cw}. In particular, 
it was shown that the electron and 
positron beam polarizations can be used to significantly enhance the signal and 
suppress the background from radiative neutrino production 
simultaneously~\cite{Choi:1999bs,Dreiner:2006sb,Dreiner:2007vm}.  
This applies in particular for scenarios where the lightest neutralino
is mainly bino, i.e., has an enhanced coupling to the right sleptons.
Right slepton exchange can be enhanced with positive electron and
negative positron beam polarizations. Since such an adjustment reduces
the couplings to the $W$ bosons, radiative neutrino production can
be severely suppressed at the same time.
In Fig.~\ref{varBeamPol},  we show the beam polarization
dependence of signal and background, for our benchmark scenario
with the relevant low energy SUSY parameters
 \begin{equation}
M_1= 103\GeV, \; M_2= 193\GeV, \; \mu= 396\GeV,  \; \tan\beta=10,\; 
m_{\tilde e_{R(L)}} \!= \!125(190)\GeV,  
\label{Eq:scenario}
\end{equation}
where the lightest neutralino $m_{\tilde\chi^0_1}= 98$~GeV 
has a bino component of $N_{11} = 0.986$.
Already for a realistic beam polarization of  $(P_{e^-},P_{e^+}) = (0.8,-0.3)$,
we obtain large signal to background ratios.  
The electroweak parameters of our scenario correspond to the 
widely studied SPS1a$^\prime$ point~\cite{AguilarSaavedra:2005pw}.
Note that although the first generation squarks and gluinos below about
$1$~TeV are excluded by recent LHC data~\cite{squarkLHC},
if their masses are roughly equal, the QCD sector of the SPS1a$^\prime$ 
point is not relevant for our study.

%

\section{Experimental analysis \label{sec:experiment}}
Due to the large number of signal and irreducible background events
in this channel, systematic uncertainties of the detector measurements
and of the beam parameters have to be considered. Therefore we perform
an experimental study~\cite{DESY-THESIS-2011-034, bartels:talk,kittel:talk}, 
to demonstrate the potential of a LC for measuring the production cross section
and the $\tilde\chi^0_1$ mass.
The analysis is made in the framework of the proposed International
Linear Collider~(ILC)~\cite{Phinney:2007zz} in full simulation of the
International Large Detector concept~(ILD)~\cite{:2010zzd}, and is valid 
for the nominal ILC parameter set as given in the 
ILC Reference Design Report~(RDR)~\cite{Phinney:2007zz}. 
The study which we shortly outline here~\cite{kittel:talk},
is a model specific application of the model independent study to search for 
WIMP Dark Matter at the ILC~\cite{bartels:talk,Birkedal:2004xn},
for further details see Ref.~\cite{DESY-THESIS-2011-034}.

\subsection{Background and event generation}

Apart from the dominant irreducible SM background process of radiative neutrino 
pair production $\ep\rightarrow\nngN$~with up to three final state photons 
($N=0,1,2$), additional SM processes have been considered:
Radiative Bhabha scattering $\ep\rightarrow \ep\gamma$
with forward peaked final state electrons,
leaving the detector through the beam pipe or hitting the forward calorimeters, as well as
multi-photon final states  $\ep \rightarrow \gamma\gamma (N)\gamma$,
where only one of the emitted photons is properly reconstructed.

\smallskip

For the event generation,
{\sc Whizard}~\cite{Kilian:2007gr,Moretti:2001zz} has been initialized
with the ILC baseline parameter set for a $\sqrt{s} = 500$~GeV machine~\cite{Phinney:2007zz}.
The production of initial state radiation~(ISR) in the 
leading logarithm approximation~\cite{Kilian:2007gr} is switched on, resulting in up to 
two predominantly soft and collinear additional photons.
The beam energy spectrum used in the event generation has been calculated
using {\sc GuineaPig}~\cite{GuineaPig}.

\subsection{Detector simulation}

For the detector simulation the {\sc Geant4}~\cite{Agostinelli:2002hh} 
based {\sc Mokka}~\cite{Mokka} simulation
software has been used. The event reconstruction
was performed with the {\sc PandoraPFA}~\cite{Thomson:2009rp} reconstruction algorithm used 
within the {\sc MarlinReco}~\cite{MarlinReco} reconstruction framework for linear colliders.  
After simulation, two corrections are performed on the reconstructed event samples.
In a first step, split uncharged electromagnetic clusters are iteratively merged
with a cone based method
to form higher level photon candidates. 
A second correction re-calibrates the detected photon energies with respect
to energy lost in the cracks and inter-module gaps for cabling 
and readout of the ILD~\cite{DESY-THESIS-2011-034}.

\begin{table}[t!]
  \centering
  \renewcommand{\arraystretch}{1.08}
  \begin{tabular*}{\textwidth}{l@{\extracolsep{\fill}} p{5mm} rr} 
 \quad                    &&      $(|\Pe|;\,|\Pp|)\,=\,(0.8;\,0.3)$  &      $(|\Pe|;\,|\Pp|)\,=\,(0.8;\,0.6)$ \\[0.5pt]
    \hline
    $\sigma_{RL}/\sigma_{0}$ &&   $3.89 \;\pm\; 0.8\;\; (0.5)\;\;\;$  &   $3.89 \;\pm\; 0.3\;\; (0.2)\;\;\;$ \\
    $\sigma_{RR}/\sigma_{0}$ &&   $0.00 \;\pm\; 1.1\;\; (0.7)\;\;\;$  &   $0.00 \;\pm\; 0.8\;\; (0.5)\;\;\;$ \\
    $\sigma_{LL}/\sigma_{0}$ &&   $0.00 \;\pm\; 1.2\;\; (1.0)\;\;\;$  &   $0.00 \;\pm\; 0.8\;\; (0.5)\;\;\;$ \\
    $\sigma_{LR}/\sigma_{0}$ &&   $0.11 \;\pm\; 1.3\;\; (0.8)\;\;\;$  &   $0.11 \;\pm\; 1.0\;\; (0.5)\;\;\;$ \\
    \hline
  \end{tabular*}
  \caption{ 
    Measured fully polarized cross sections 
    $\sigma_{\lbrace R,L\rbrace}  /\sigma_0 \,\pm\,\delta_{\sigma}$~[fb], see Eq.~(\ref{Eqn:Deconstruct1}),
   for two different positron polarizations for scenario Eq.~(\ref{Eq:scenario}).
   The error $\delta_{\sigma}$ is the squared sum of the statistical and the systematic errors.
    The values in brackets correspond to an improved polarization measurement of $\delta P = 0.1\%$,
    instead of $\delta P = 0.25\%$.  
}
  \label{Table:NeutXSecPolMeas1}
\end{table}

\subsection{Event sampling and selection cuts}

In order to enable an efficient study of many different models,
 the signal events were generated by reweighting the 
fully simulated SM background samples. To avoid statistical correlation,
the background samples have been divided into three statistically  independent 
subsamples. The first two subsamples are used for the  background and signal 
contribution to the data.  The third sample is parametrized to yield 
the predicted photon spectra. 

\medskip

Since the $\nng$ SM background is indistinguishable on an event-by-event basis
from the neutralino signal, the selection is tuned to isolate this SM background.
An event is considered signal-like when it contains at least one photon with 
 \begin{equation}
10\GeV < E_{\gamma} < 220\GeV,
\qquad
 |\cos{\theta_\gamma}| < 0.98.
\label{Eq:cuts}
\end{equation}
This signal definition ensures that the detected
photons are within the tracking acceptance of the ILD detector
to distinguish them from charged particles. The cut on the photon energy
reduces the contributions from soft ISR, and excludes the massless neutrino 
final states from the radiative $Z$-return at photon energies of $241$~GeV,
see  Fig.~\ref{plotEdist}.

\smallskip

In the further event selection, the following  constraints are set to reject
the dominant reducible SM backgrounds. 
To exclude hadronic and leptonic final states,  a cut
$E_{\rm vis}-E_{\gamma}<20$~GeV is applied on the maximal exclusive energy,
i.e., the total detected (visible) event energy excluding the selected photon.
The maximal transverse track momentum  is constrained to $p_{T} < 3$~GeV.
Low $p_{T}$ tracks have to be allowed because of 
track overlays from $\ep$ pairs from the beamstrahlung background,
and from multi-peripheral $\gamma\gamma\rightarrow$~hadrons events.
Although the angular distribution of the final state fermions is strongly peaked in the
forward direction, a strong activity in the sensitive volume is expected,
due to the large cross section in the order of $5\times10^{8}$~fb.
The impact of both contributions has been studied for the ILD detector
using {\sc Whizard} for event generation~\cite{:2010zzd}. The beamstrahlung spectrum
has been simulated with {\sc Guinea Pig}.  
On average $0.7$ tracks from $\gamma\gamma$ processes and $1.5$ tracks
from beamstrahlung background are expected per bunch crossing. Tight selection
criteria on these tracks would therefore reduce the signal statistic strongly.
Finally, the large Bhabha background is reduced by an identification of
high energy electrons in the forward beam calorimetry.
The selection efficiency of the $\nng$ background is above $80\%$ on average, 
while the multi-photon and Bhabha background is reduced to $<1\%$.
The signal selection efficiency is about $92\%$.

\subsection{Cross section measurement and coupling structure}

The signal cross section is determined
by a subtraction of the expected number of background events  $<N_{B}> $
from the number of observed data events  $N_{D}$,
\begin{eqnarray}\label{Eqn:SigmaMeas}
\sigma(\Pe,\Pp) & = &\frac{N_{D}-<N_{B}>}{\lum \times \varepsilon},
\end{eqnarray}
with the experimental luminosity $\lum$ and the signal efficiency $\varepsilon$.
With four different polarization configurations, the fully polarized cross sections 
$\sigma_{\lbrace L,R \rbrace}$, and hence the helicity structure 
of the coupling to the beam electrons, can be  determined from~\cite{MoortgatPick:2005cw}
\begin{eqnarray}
 \sigma( P_{e^-}, P_{e^+}) & = &
\frac{1}{4}\Big[ (1+P_{e^-})(1+P_{e^+}) \sigma_{RR} 
                  + (1-P_{e^-})(1-P_{e^+}) \sigma_{LL}\nonumber\\
        && + \,\,\, (1+P_{e^-})(1-P_{e^+}) \sigma_{RL} 
                  + (1-P_{e^-})(1+P_{e^+}) \sigma_{LR} \Big].
\label{Eqn:Deconstruct1}
\end{eqnarray}
In Tab.~\ref{Table:NeutXSecPolMeas1}, we summarize the reconstructed cross sections
and their corresponding errors.
The luminosity of $\lum = 500\fb^{-1}$ is distributed to 
the odd (even) sign polarization configurations with $\lum = 200\,(50)\fb^{-1}$ each.
With an assumed  systematic polarization measurement error of 
 $\delta P = 0.25\%$~\cite{Boogert:2009ir},
the cross section $\sigma_{RL}$ can be determined to a level of about $20\,(7)\%$,
for an positron polarization of $|\Pp|= 0.3\,(0.6)$.
Since the systematic uncertainty is dominated by the precision of the
polarization measurement $\delta P$,
the systematic error can be reduced by about a third if $\delta P = 0.1\%$
could be achieved, see~Tab.~\ref{Table:NeutXSecPolMeas1}.
At the ILC, the long term average of the luminosity weighted beam polarization 
can be determined from collision data to $0.17\%$ for both electrons and
positrons with $|\Pp| = 0.6$, dominated by the uncertainties from the polarimeters~\cite{Marchesini}. 
Under the assumption of uncorrelated polarization errors,
a combination of the measured fully polarised cross sections $\sigma_{\lbrace L,R \rbrace}$
yields an uncertainty of $9.3\,(6.0)\%$ on the cross section for unpolarised
beams for $\delta P = 0.25\%$ and $|\Pp|= 0.3\,(0.6)$, again systematically dominated by the polarimeters. 


\subsection{Neutralino mass measurement}

The mass of the neutralino candidate is determined from a $\chi^{2}$  fit 
of the full data energy spectrum to template spectra with different masses.
Depending on the degree of positron polarization, statistical precisions of
1.7 to 2.6 GeV can be obtained for an integrated luminosity
of $500\fb^{-1}$. The total uncertainty on the candidate mass 
is dominated by the uncertainty in the beam energy spectrum,
contributing with an error of $2.2$~GeV to the mass determination, see Tab.~\ref{Table:NeutMassMeasuremnt}.
The influence of the beam energy spectrum, which distorts the energy spectrum of the signal photon, 
has been estimated by generation of
signal spectra in a generic WIMP model~\cite{Birkedal:2004xn} for two
different sets of beam parameters, RDR~\cite{Phinney:2007zz}  
and SB-2009~\cite{Berggren:2010wy}. 
This is a conservative estimate, since the beam energy spectrum 
will be known to a higher degree than the difference between the sets.

\begin{table}[t]
  \centering
  \renewcommand{\arraystretch}{1.10}
  \begin{tabular*}{0.80\textwidth}{c@{\extracolsep{\fill}} p{4mm} rr}  
    \multicolumn{4}{c}{\quad } \\[-4.8mm]
   $m_{\nt_{1}}$ [GeV] && $\pm\,$stat. $\pm\,$sys.$\,(\delta E \pm \delta\lum)\;$ (total)$\;$ [GeV] & 
               $(\Pe;\,\Pp)$ \\[0.5pt]
    \hline
 $\,97.7$  && $\pm\,2.65 \;\pm\,0.09 \;\pm\,2.20\;\;\; (3.44)\;\;\;$  &  $(0.8;\, 0.0)$ \\
 $\,97.7$  && $\pm\,2.07 \;\pm\,0.09 \;\pm\,2.20\;\;\; (3.02)\;\;\;$  &  $(0.8;\,-0.3)$ \\
 $\,97.7$  && $\pm\,1.70 \;\pm\,0.09 \;\pm\,2.20\;\;\; (2.79)\;\;\;$  &  $(0.8;\,-0.6)$ \\[0.5pt]
    \hline 
  \end{tabular*}
  \caption{
    Neutralino mass determined from a template comparison for an integrated 
    luminosity of $\lum = 500\fb^{-1}$ and different beam polarizations. 
    The systematic uncertainties comprise of the beam energy scale calibration ($\delta E$) 
    and the beam energy spectrum ($ \delta\lum$).
}
  \label{Table:NeutMassMeasuremnt}
\end{table}


\newpage

\begin{footnotesize}


\end{footnotesize}



\begin{thebibliography}{99}

\bibitem{Haber:1984rc}
  H.~E.~Haber and G.~L.~Kane,
  Phys.\ Rept.\  {\bf 117}, 75 (1985).


\bibitem{AguilarSaavedra:2005pw}
  J.~A.~Aguilar-Saavedra {\it et al.},
  Eur.\ Phys.\ J.\  C {\bf 46}, 43 (2006)
  [arXiv:hep-ph/0511344].


\bibitem{Choi:1999bs}
  S.~Y.~Choi, J.~S.~Shim, H.~S.~Song, J.~Song and C.~Yu,
  Phys.\ Rev.\ D {\bf 60} (1999) 013007
  [arXiv:hep-ph/9901368].

\bibitem{Baer:2001ia}
  H.~Baer and A.~Belyaev,
  arXiv:hep-ph/0111017.

\bibitem{Dreiner:2006sb}
  H.~K.~Dreiner, O.~Kittel and U.~Langenfeld,
  Phys.\ Rev.\  D {\bf 74} (2006) 115010
  [arXiv:hep-ph/0610020].

\bibitem{Dreiner:2007vm}
  H.~K.~Dreiner, O.~Kittel and U.~Langenfeld,
  Eur.\ Phys.\ J.\  C {\bf 54}, 277 (2008)
  [arXiv:hep-ph/0703009].

\bibitem{Dreiner:2009ic}
  H.~K.~Dreiner, S.~Heinemeyer, O.~Kittel, U.~Langenfeld, A.~M.~Weber, G.~Weiglein,
  Eur.\ Phys.\ J.\  {\bf C62}, 547-572 (2009).
  [arXiv:0901.3485 [hep-ph]].

\bibitem{MoortgatPick:2005cw}
  G.~Moortgat-Pick {\it et al.},
  Phys.\ Rept.\  {\bf 460}, 131 (2008)
  [arXiv:hep-ph/0507011].


\bibitem{squarkLHC}
  G.~Aad {\it et al.}  [ATLAS Collaboration],
  Phys.\ Rev.\ D {\bf 85}, 012006 (2012)
  [arXiv:1109.6606 [hep-ex]];
%
  S.~Chatrchyan {\it et al.}  [CMS Collaboration],
  Phys.\ Rev.\ Lett.\  {\bf 107}, 221804 (2011)
  [arXiv:1109.2352 [hep-ex]].

\bibitem{DESY-THESIS-2011-034}
  C.~Bartels, Diploma Thesis, 
  \emph{Model-independent WIMP searches at the ILC}, DESY 2007;
%
   C.~Bartels, PhD Thesis, 
   \emph{WIMP search and a Cherenkov detector prototype for ILC polarimetry},
   DESY-THESIS-2011-034.

\bibitem{kittel:talk}
\verb$http://ilcagenda.linearcollider.org/contributionDisplay.py?contribId=263&sessionId=6&confId=5134$

\bibitem{bartels:talk}
\verb$http://ilcagenda.linearcollider.org/contributionDisplay.py?contribId=257&sessionId=6&confId=5134$

 \bibitem{Phinney:2007zz}
   N.~Phinney,
   ICFA Beam Dyn.\ Newslett.\  {\bf 42} (2007) 7.

 \bibitem{:2010zzd}
   T.~Abe {\it et al.}  [ILD Concept Group - Linear Collider Collaboration],
   arXiv:1006.3396 [hep-ex].

\bibitem{Birkedal:2004xn}
  A.~Birkedal, K.~Matchev and M.~Perelstein,
  Phys.\ Rev.\ D {\bf 70} (2004) 077701
  [arXiv:hep-ph/0403004].

 \bibitem{Kilian:2007gr}
   W.~Kilian, T.~Ohl and J.~Reuter,
   Eur.\ Phys.\ J.\  C {\bf 71} (2011) 1742
   [arXiv:0708.4233 [hep-ph]].

 \bibitem{Moretti:2001zz}
   M.~Moretti, T.~Ohl and J.~Reuter,
   arXiv:hep-ph/0102195.
 
 \bibitem{GuineaPig}
    D.~Schulte,
    TESLA 97-08 (1996).
 
\bibitem{Agostinelli:2002hh}
  S.~Agostinelli {\it et al.}  [GEANT4 Collaboration],
  Nucl.\ Instrum.\ Meth.\  A {\bf 506}, 250 (2003).
 
 \bibitem{Mokka}
  {\sc Mokka}, \verb$http://polzope.in2p3.fr:8081/MOKKA$
 
 \bibitem{Thomson:2009rp}
   M.~A.~Thomson,
   Nucl.\ Instrum.\ Meth.\  A {\bf 611} (2009) 25
   [arXiv:0907.3577 [physics.ins-det]].
 
\bibitem{MarlinReco}
   {\sc MarlinReco},
    \verb$http://ilcsoft.desy.de/portal/software\_packages/marlinreco$

\bibitem{Boogert:2009ir}
  S.~Boogert  {\it et al.},
  JINST {\bf 4} (2009) P10015
  [arXiv:0904.0122 [physics.ins-det]].

 \bibitem{Marchesini}
   I.~Marchesini,
     DESY-THESIS-2011-044.

\bibitem{Berggren:2010wy}
  M.~Berggren,
  arXiv:1007.3019 [hep-ex].











\end{thebibliography}
\end{document}